\documentclass[seceq]{ptptex}

\usepackage{graphicx}
\usepackage{wrapft}

\title{
Correlations of Power-law Spectral and QPO Features In Black Hole Candidate
Sources%
}

\author{
Ralph \textsc{Fiorito}$^{1}$
and Lev \textsc{Titarchuk}$^{2}$ 
}

\inst{
$^1$ Catholic University of America and NASA/GSFC\\
$^2$ George Mason University/Naval Research Laboratory\\
}

\abst{
Recent studies have shown that strong correlations are observed between
low frequency QPO`s and the spectral power law index
for a number of black hole candidate sources (BHC's), when these sources exhibit quasi-steady hard x-ray emission states. 
The dominant long standing interpretation of QPO's is that
they are produced in and are the signature of the thermal accretion
disk. Paradoxically, strong QPO`s are present even in the cases where the thermal component is negligible. 
We present a model which identifies
the origin of the QPO's and relates them directly to the properties of a
compact coronal region which is bounded by the adjustment from Keplerian 
to sub-Kelperian inflow into the BH and which is primarily responsible for the observed power law spectrum. The model also predicts the relationship between high and
low frequency QPO's and shows how BH's can be unique identified
from observations of the soft states of NS's and BHC's. 
}

\begin{document}

\maketitle

\section{Introduction}
A number of observations of BHC's do not show consistent correlations of low QPO frequency with disk
parameters [1,2]. On the other hand, strong consistent correlations
between power law spectral index and low frequency QPO's have been recently observed [3]. In addition, there is mounting observational evidence that the 
large number of  spectral "states" formerly developed by
to explain the wide variability of BHC's such as
GRS1915+105, can be reduced to a few canonical states, i.e. a hard
state with spectral power law index
 $\Gamma \sim1.6\pm 0.1$ , a soft or "extended" power-law state
characterized by $\Gamma \sim 2.7\pm0.2$, and a thermal state [4].
\par
  These data have prompted us to introduce a model, i.e. the
Transition Layer (TL) model [5] to explain the correlations observed. The
main feature of the TL model is a hot compact region near the BH which serves as
the primary region for Compton upscattering of soft disk photons.
The TL model shows how the QPO's are related to the
size, optical depth, temperature
and spectral index and predicts the correlation between
index and QPO frequency.
\section{Predictions of the Model}


The model predicts
two generic classes or states for accreting BH's:
\par
{\it HARD STATE}:  
In this state the energy release in the
outer boundary of the corona $Q_{cor}$ 
is much greater than the energy release in
the disk $Q_d$  ($Q_{cor}\gg  Q_d$).
Also: a) the mass accretion rate in the disk $\dot m=\dot M/\dot M_{\rm Edd}$  
is small,
b) the optical depth of the corona is order of unity  and at least few
times larger than the mass accretion rate in the disk $\dot m$,
c) 
the Compton $y$ parameter in the hard state is almost a universal constant and is
independent of $Q_d$ ; this leads to a universal value of the photon index $\Gamma\sim 1.6\pm
0.1$, and
d) the efficiency for photon upscattering is
 second order in  $V/c$, where V is the mean plasma thermal velocity
(thermal Comptonization regime).
\par
{\it SOFT STATE}:  In this state the opposite condition applies, i.e. 
 $Q_{cor} << Q_d$. 
As the
mass accretion rate increases the system goes to soft state and the photon
index saturates to an asymptotic value  $\Gamma = 2.7\pm0.2$ 
 depending on the temperature of the flow, which is of the order of the photon disk temperature 
 (1-10 keV). The efficiency for upscattering is determined by
first order in $V/c$   ( bulk inflow Comptonization regime).
\par
The QPO low  frequency $\nu_{low}$   is associated with 
the magnetoacoustic oscillation  of the transition layer (cavity)
i.e. $\nu_{low}\sim V/L$ ; where $L$ is a cavity size. The QPO high frequency $\nu_{high}$  is related to the Keplerian frequency at
the outer TL boundary radius.

\section{Data Interpretation and Summary}
\begin{wrapfigure}{r}{6.6cm} 
\centerline{\includegraphics[width=50 mm,height=8 cm, angle=-90]
                         {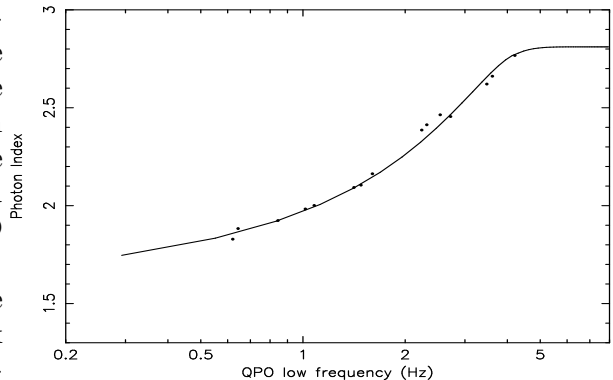}}
\caption{Plot of power law photon index versus QPO centroid
frequency for the plateau observations of GRS1915+105 from [3] along with a fit using the TL Model with $M =12 M_{\odot}$, $\tau_0=\gamma^{1.25}$.
}
\label{fig:2}
\end{wrapfigure}
 A fit to the data (Fig.1) is obtained from analytical
relationships [6] associating: 1) the Reynolds number $\gamma$ for the accretion flow, which
is a function of the mass accretion rate and the viscosity in the disk,
and the size of the TL;  
 2) the optical depth, plasma
temperature and spectral index of the TL; and 3)the QPO
frequency and spectral index. \par
In contrast to BH's, NS's in the high/soft
state do not show high index power-law spectra, but rather a
blackbody like spectrum due to the presence of the surface and radiation pressure. 
Bulk inflow Comptonization present in the case of a BH is never present for
NS's. Thus, the observed saturation of the QPO frequency with spectral indices 
of $\Gamma\sim 2.7$ is a unique signature of BH's and can be used to identify them. 

 

%

\end{document}